\documentclass[journal]{IEEEtran}

\newif\ifarxiv
\arxivtrue

\usepackage[utf8]{inputenc}
\usepackage{amsmath}
\usepackage{graphicx}
\graphicspath{{figures/}}
\usepackage{booktabs}
\usepackage[numbers, sort&compress]{natbib}
\usepackage{doi}

\ifarxiv
  \newcommand{\SuppRef}[1]{\ref{#1}} 
\else
  \usepackage{xr-hyper}  
  \newcommand{\SuppRef}[1]{\ref{SM-#1}}
\fi

\usepackage{hyperref}
\usepackage{xurl}

\newcommand{\MainRef}[1]{\ref{#1}}

\hypersetup{
    colorlinks=true,
    linkcolor=blue,
    filecolor=magenta,
    urlcolor=cyan,
    citecolor=blue,
}

\title{Slower but Faster: When Self-Organized Traffic Flows Beat Traffic Signal Control}
\date{\today}

\begin{document}

\author{Damian~Dailisan,
        Meisam~Akbarzadeh,
        and~Dirk~Helbing%
\thanks{D.~Dailisan and D.~Helbing are with Computational Social Science, ETH Z{\"u}rich, Z{\"u}rich, Switzerland (e-mail: ddailisan@ethz.ch). Corresponding author: D.~Dailisan.}%
\thanks{M.~Akbarzadeh is with the Department of Transportation Engineering, Isfahan University of Technology, Isfahan, Iran.}%
\thanks{D.~Helbing is also with the Complexity Science Hub, Vienna, Austria.}}

\maketitle






\begin{abstract}
Urban traffic flow networks are complex adaptive systems, whose variability challenges centralized, top-down control. We study decentralized alternatives in which order emerges from local interactions, comparing the network-wide performance of roundabouts in SUMO against fixed-time and actuated control as well as self-organizing signals on a 4$\times$4 grid and a rewired, irregular variant of it, across demand levels and periodic surge amplitudes. In our computer simulations of the regular grid, converting the single most central intersection already lowers delays by 2 to 3\% relative to actuated control; each additional conversion lowers them further, and full conversion outperforms every signalized alternative under all conditions studied.
In the irregular traffic flow network with full conversion of all intersections into roundabouts, this holds only up to a critical load, where traffic flow
breaks down sharply, when the time-averaged demand reaches about 3.1 times the base load.
Beyond that point, the self-organizing traffic signals show the lowest delay of any network-wide strategy and is nearly insensitive to surges.
Interestingly, roundabouts are favorable solutions even at lower speed limits: in the regular grid, compared to the mean speed under actuated control at a speed limit of 50\,km/h, roundabouts match that speed at a speed limit of about 40\,km/h.
Compared to traffic signal control, this establishes a slower-is-faster effect. 
Furthermore, travel times in self-organized traffic flows are more predictable than in networks with conventional signals.
\end{abstract}

\begin{IEEEkeywords}
Roundabouts, traffic signal control, self-organization, decentralized traffic management, network resilience, microscopic traffic simulation.
\end{IEEEkeywords}

\section{Introduction}

Urban traffic flows are a complex adaptive system in which local interactions give rise to traffic jams, stop-and-go waves, and network-wide congestion \cite{Helbing2001}. This makes it hard to predict and manage traffic flows \cite{timotheou2013towards}, which is an NP-hard optimization problem.
The prevailing solution is centralized traffic signal control, which pre-calculates timings from historical or predicted demand \cite{Hunt1982SCOOT, Aboudolas2009StoreForward}.
Yet, the ``typical conditions'' for which such plans are optimized rarely occur \cite{wei2019survey}: demand fluctuates from minute to minute, such that deviations from the calculated signal plans cause unnecessary delays and stops \cite{Day2010Retiming}. 
While online adjustment of offsets and splits try to fix the mismatch \cite{Abbas2001Offset}, even adaptive systems have to rely on simplified signal control approaches for large networks \cite{Aboudolas2009StoreForward}. 
The behavior of such networks under surges and disruptions is a central focus of resilience research \cite{Mattsson2015Resilience, Korecki2023Disruptions}.

\begin{figure}[tb]
    \centering
    \includegraphics{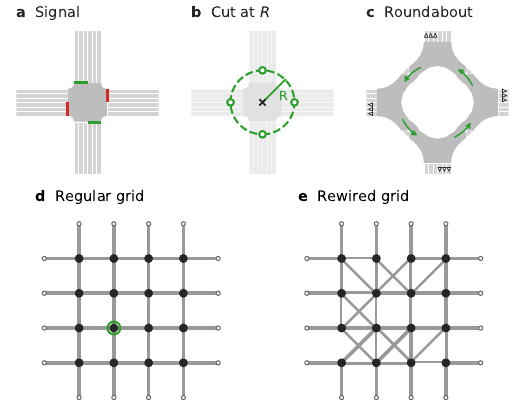}
    \caption{(a)--(c) Conversion of the intersection circled in (d): (a) signalized, where colored stop bars show a signal phase; (b) the traffic signal is removed and approaches cut at radius $R = 22.5$\,m, with a priority junction at each cut; (c) connection of these junctions using ring edges. At each junction, incoming vehicles yield (triangles).
    (d) Regular 4$\times$4 grid and (e) rewired grid; filled circles represent intersections, open circles network entries and exits.}
    \label{fig:conversion}
\end{figure}

Decentralization and self-organization offer an alternative, in which robust global order emerges from local interactions without a central authority \cite{Helbing2005SelfOrganized, Laemmer2008SelfControl, gershenson2015selforganizing}. This has inspired novel kinds of software for
signals that switch based on local, real-time information. Examples reach from self-organizing priority rules \cite{Laemmer2008SelfControl} to max-pressure control \cite{Varaiya2013MaxPressure} and multi-agent reinforcement learning \cite{Chu2020MARL, Haydari2022Survey}. The underlying decisions of these approaches can be decentralized, and one could even allow road users to vote for one of several signal control policies, each trained for a different goal. Such democratic approaches to governing intersections can achieve high performance with respect to several (``pluralistic'') goals \cite{Korecki2024Democratizing}.
Focusing now on hardware, roundabouts pursue the same idea of self-organization, but with a single spatial priority rule, yielding circular traffic flows (Fig.~\ref{fig:conversion}) \cite{Rodegerdts2010NCHRP672, Persaud2001Safety}.
The comparison made here is, therefore, between two forms of decentralization and self-organization: (i) a \textit{spatial} rule embedded in junction geometry and (ii) a \textit{temporal} rule implemented in the signal software, such as the Analytic+ implementation \citep{Korecki2022Analytically} of the self-control approach proposed by \citet{Laemmer2008SelfControl}.

At a single junction, the case for roundabouts is largely settled, with a well-established safety benefit \cite{Elvik2017Meta, Persaud2001Safety} and capacity models supporting routine design \cite{Rodegerdts2010NCHRP672}. In contrast, their suitability for road networks is still disputed.
In Switzerland, many roundabouts introduced since the mid-1980s now saturate at rush-hour peaks, prompting a shift back to signals: Lucerne plans to replace seven roundabouts, Ostermundigen has replaced three with sensor-controlled signals, and the Federal Roads Office reports simulations favoring signals on a dense section of the A13, though opponents argue roundabouts work well off-peak \cite{Aschwanden2026NZZ}.
The literature offers little guidance, because roundabouts have been mostly studied one junction at a time. 
The few corridor and grid studies \cite{Brocchini2025Paradoxes, Taglieri2023NetworkWide}, however, consider neither self-organizing signals, demand that varies within the operating period, nor partial conversion (Sec.~\ref{sec:related}), which are precisely the conditions of the Swiss case.
Gap-acceptance models take the arrival flows at each entry as given and, therefore, cannot represent junctions that spill back into one another \cite{Daganzo2007Gridlock, Akcelik2011Metering}, while network studies of decentralized signal control treat junction layouts as fixed \cite{Noaeen2022Review}.

We address these gaps with SUMO simulations of a regular 4$\times$4 grid and a rewired, irregular variant of it, comparing full and partial conversions of intersections into roundabouts with fixed-time and actuated traffic signal control as well as two kinds of self-organizing signals under periodically varying demand surges. We ask three research questions:
\begin{enumerate}
    \item At what demands and topological conditions do roundabout networks outperform networks with signalized intersections?
    \item How can a partial conversion of intersections deliver a proportionate share of the benefit?
    \item Can roundabout networks with reduced speed limits produce faster average speeds than signalized networks at higher speed limits, thereby demonstrating a ``slower-is-faster'' effect \cite{Gershenson2015SlowerIsFaster}?
\end{enumerate}

\section{Related work}
\label{sec:related}

\subsection{Comparative Performance of Intersection Designs}

Safety gains are the least contested advantage of the roundabouts.
Roundabouts reduce the possible crossing paths between streams of vehicles, so the remaining conflicts occur at low speeds and shallow angles \cite{Rodegerdts2010NCHRP672, Persaud2001Safety}.
Conversions of signalized intersections into roundabouts achieve about 35\% fewer total collisions; this means up to 80\% fewer injuries and 90\% fewer fatal collisions \cite{Rodegerdts2010NCHRP672, Persaud2001Safety}.

Operational performance is more disputed, as it depends on traffic conditions.  
Under low to moderate volumes, conversions of signalized intersections into roundabouts reduce delays by 20\% to 89\% and stops by 56\% \cite{Rodegerdts2010NCHRP672}. Capacity, however, is determined by gap acceptance in the circulating stream \cite{Guo2019GapAcceptance}: as volumes rise, gaps shrink and queues form \cite{Vlahos2008Capacity}, particularly when one stream dominates the minor approaches \cite{Persaud2001Safety}. Beyond a certain threshold, a signalized intersection with dedicated turning lanes can serve more vehicles \cite{Vlahos2008Capacity, Fortuijn2009Turbo}.
Practitioners respond to dominant circulating streams with metering signals \cite{Akcelik2011Metering}, which shows that a spatial priority rule alone cannot break a deadlock. While spillover effects of a saturated roundabout in a network can block neighboring intersections or roundabouts, signals are believed to retain the ability to clear queues with their phasing.

However, network-level evidence on this question is thin, and most comparisons focus on a single junction \cite{Zhou2022}. In Pisa, converting a signalized corridor into a sequence of roundabouts unexpectedly increased congestion, which is considered to be a network paradox~\cite{Brocchini2025Paradoxes}.
Similarly to our study, \citet{Taglieri2023NetworkWide}, compares grids with roundabouts vs. signals and finds that single-lane roundabout networks outperform signals in terms of free-flow speed and flow capacity, while two-lane networks do not. However, it neither considers adaptive signals, nor partial conversions, nor time-varying demand.

Environmental effects also depend on the traffic state: fewer stops lower emissions after conversion \cite{Mandavilli2008Emissions}, mainly below a demand-to-capacity ratio of about 0.7 \cite{Salamati2015Emissions}, but more frequent acceleration can offset this improvement \cite{Hallmark2011OnRoad, Varhelyi2002SmallRoundabouts}.

\subsection{Decentralized, Self-Organizing Approaches}

Traditionally, a control center manages traffic signals to coordinate citywide traffic flows in space and time.
The oldest approach is based on fixed-time control, which tries to achieve a synchronization of periodic intersection signal schemes \cite{Gartner2002Arterial}.
Over time, cities have switched to adaptive signals, which adjust green times, offets, and splits to fluctuations in the traffic flows \cite{wei2019survey, Hunt1982SCOOT}. 

Recently, it is often proposed to use signal control approaches based on machine learning, typically reinforcement learning (RL) \cite{Chu2020MARL, Haydari2022Survey}. 
Within the framework multi-agent simulations (MARL), this has been even applied to thousands of intersections \cite{Chen2020ThousandLights}. 
However, the approach degrades under disruptions and out-of-distribution demand \cite{Korecki2023Disruptions}, while heterogeneous evaluation protocols make it difficult to compare the actual performance across studies \cite{Noaeen2022Review}. 

A further approach to traffic signal control was inspired by the fact that traffic can be modeled as a self-driven many-particle system, in which self-organized spatio-temporal patterns (such as stop-and-go waves) emerge from local interactions, even without central coordination \cite{Helbing2001, Helbing2005SelfOrganized}. 
Applying such self-organization principles, \citet{Laemmer2008SelfControl} have proposed an innovative self-control approach. 
Here, the traffic flows drive the traffic signals rather than the other way round. 
This is based on a short-term anticipation of vehicle arrivals and switching phases such that the number of stopped vehicles is minimized. 
Within this framework, coordinated green waves emerge as a result of self-organization. 

It has been shown that self-organizing signals of this kind can outperform parameter-heavy reinforcement learning~\cite{Korecki2023Disruptions,Bompard2026RuleBased}. Therefore, in our paper, we use two simple, rule-based signals, Analytic+ and the rule-based AcoLight, as performance benchmarks.
Complementary, we would like to mention max-pressure control \cite{Varaiya2013MaxPressure}, which can stabilize network queues based on local information. It has been adapted to deployed hardware \cite{Levin2020MaxPressure} and tested in the field \cite{Mercader2020MaxPressure}. Successful implementations also exist for variants of the Analytic+ approach \cite{selfcontroltechreport2020}.

All of these approaches allow one to decentralize switching schedules, but keep the junction geometry fixed. In contrast, turning signalized intersections into roundabouts removes switching schedules, replacing them by a priority rule, where moving decisions of driver-vehicle units are only based on local information \cite{Noaeen2021} (basically the availability of a large enough gap in the traffic flow).

\subsection{Network-Level Performance and Structure}

At the scale of a whole network, traffic performance is commonly summarized by the macroscopic fundamental diagram (MFD), which relates the average flow in a network to its average density \cite{Geroliminis2008MFD, Daganzo2008MFD, helbing2009derivation}.
Flow rises with the number of vehicles in the network up to a critical accumulation and declines beyond it as congestion spreads.
This critical point differs strongly between cities and is largely explained by the topology of their road and bus networks \cite{Loder2019Capacity}.
Within a network, the same average density can yield lower flows when vehicles are distributed unevenly \cite{Mazloumian2010Spatial}, and gridlock can arise from instability in the accumulation dynamics rather than from high demand alone \cite{Daganzo2007Gridlock}.

The structure of a road network also shapes how it performs and how it absorbs disruptions \cite{Akbardazeh2018Communicability, Ganin2017Resilience}.
Resilience studies therefore rank links or nodes by the network-wide cost of their degradation \cite{Mattsson2015Resilience, Zhou2019Resilience}.
A common structural ranking is betweenness centrality, which counts how often a node lies on shortest paths, but it correlates only moderately with observed flows \cite{Gao2013Betweenness}.
In network traffic models with shortest-path routing, however, the load on each node scales with its betweenness, so the most central nodes determine the onset of congestion \cite{Zhao2005Onset}.
Weighted measures such as node strength, the summed weight of a node's incident edges, take the traffic into account \cite{Barrat2004Weighted}.

\section{Simulation Framework and Experimental Design}
\label{sec:methods}

All simulation experiments were designed and executed using the open-source microscopic traffic simulation package SUMO (Simulation of Urban MObility) \texttt{v1.27.0}\cite{SUMO2024}.

\subsection{The SUMO Simulation Environment}

All our simulations used a homogeneous fleet of vehicles with the default car-following~\cite{Krauss1997} and lane-changing~\cite{Erdmann2015} models of SUMO.

\subsubsection{Network Topology}

We studied two synthetic networks (Fig.~\ref{fig:conversion}d,e).
The first was a homogeneous 4$\times$4 grid from the RESCO benchmark~\cite{Ault2021RESCO}, consisting of 16 intersections (nodes) and 80 directed edges in a regular, Manhattan-style layout.
The second was obtained by rewiring a random share of the grid's edges to diagonal neighbors, adding new street directions and non-four-way intersections.
Real cities vary widely in both respects~\cite{Boeing2019Order}, and topology largely determines urban traffic capacity~\cite{Loder2019Capacity}.
The rewired grid therefore tested whether results carry over to a less regular network.
Intersections were spaced 300\,m apart, and both networks have three-lane roads with a baseline speed limit of 50 km/h.

\subsubsection{Traffic Demand Modeling}

The networks were subject to a base demand of 1472 vehicles per hour, with the same origin--destination demand pairs applied to both networks.
To preserve identical route demand across traffic management strategies and avoid performance differences due to route choice, we disabled vehicle rerouting.
We tested each traffic management strategy under increasingly saturated conditions by scaling base demand with a factor $\lambda \in \{1, 2, 3\}$ as well as intermediate values $\lambda \in \{2.25, 2.5, 2.75\}$ in the rewired network to more accurately determine the demand at which the fully converted network breaks down.
On top of each demand level, we superimposed periodic demand surges with vehicle insertion rates alternating every 600 seconds (10 minutes) between $\lambda$ and a surged value of $(\lambda + a)$ times the base demand.
The surge amplitude $a$ was varied from 0 (steady demand) to 1.
Each simulation covered 4 hours (14400 seconds) of traffic, corresponding to 12 complete surge--recovery cycles.

\subsection{Signalized Intersections}

We compared five traffic management strategies: roundabouts, fixed-time and actuated control, and two kinds of self-organizing signals that decide from local measurements.

Roundabouts were generated algorithmically from the original intersections (Fig.~\ref{fig:conversion}).
A circle of radius 22.5\,m (recommended for multi-lane roundabouts~\cite{Overkamp2009Roundabouts}) was centered on the intersection; every approach road was cut where it crossed the circle, and a priority junction was placed at each cut.
Arc-shaped ring edges joined consecutive junctions with the lane count of the approaches (three lanes).
The roundabout inherited the speed limit of the main roads.
In lieu of traffic signals, entering vehicles just yielded to circulating traffic.

Fixed-time control was the non-adaptive baseline, with a default 90\,s signal plan generated for each intersection by SUMO \cite{SUMO2024}.
On the regular grid, each cycle consisted of two 33\,s phases serving through-traffic and permitted left turns, followed by a 6\,s protected left-turn phase, with 3\,s yellow intervals between phases.
Actuated control had similar phase timings, but it could extend a green phase up to a maximum duration as long as vehicles keep arriving.
It adapted to local fluctuations in demand, but had no awareness of the network.

The third strategy was Analytic+~\cite{Korecki2022Analytically}, a self-organizing signal method that combines an optimization rule with a stabilization rule~\cite{Laemmer2008SelfControl}.
Under the optimization rule, Analytic+ served the traffic flow with the highest priority score
\begin{equation}
    \pi_i=\frac{\hat{n}_i}{\tau_{i}^\mathrm{pen}+\tau_i+\hat{g}_i},
    \label{eq:priority}
\end{equation}
where $\hat{g}_i$ is the green time required to clear movement $i$, $\tau_i$ is the penalty for switching to another traffic flow, $\tau_{i}^\mathrm{pen}$ is the penalty for switching back, and $\hat{n}_i$ is the number of vehicles expected to be served in the period $\tau_{i} + \hat{g}_i$.
This rule minimizes the total waiting time of vehicles at the intersection.
The stabilization rule overrode it when this was needed to prevent excessive waiting times or spillover effects, and it was parametrized by the desired average waiting time $T_\mathrm{avg}$ and the maximum waiting time $T_\mathrm{max}$ of vehicles at the intersection.
We set $T_\mathrm{avg} = 180$\,s and $T_\mathrm{max} = 240$\,s on both networks.
The stabilization rule was triggered when the number of waiting vehicles exceeded the critical threshold
\begin{equation}
    n_i^\mathrm{crit}=Q_i T_\mathrm{avg}  \frac{T_\mathrm{max}-z_i}{T_\mathrm{max}-T_\mathrm{avg}},
\end{equation}
where $Q_i$ is the average arrival rate of flow $i$ and $z_i$ is the variable service interval between two successive service processes of flow $i$. The flow exceeding its threshold was then served next.

Because Analytic+ could switch between \textit{any} two phases rather than only to the next phase 
of a fixed-time or adaptive program, every switch was preceded by a synthesized 3\,s yellow interval.
A 2\,s all-red interval followed the yellow on every transition that ended a permitted movement that the next phase did not serve (e.g. an unprotected left turn). Therefore, vehicles inside the junction waiting to complete a permitted turn could leave before conflicting traffic is released.
The fixed-time and actuated programs needed no such interval, since their fixed phase sequence always followed a permitted phase with the protected turning phase that clears those vehicles.

The fourth strategy was AcoLight~\cite{bompard2025}, a self-organizing phase scheduler built from simple rules. 
It is one of the simple signal methods that outperformed reinforcement-learning methods in a recent benchmark study~\cite{Bompard2026RuleBased}.
Each green phase was monitored by lane-area detectors covering the last 34\,m before the stop line and the first 20\,m after the upstream junction.
A phase was held until it reached a maximum green time (90\,s), until no vehicle was detected on its entry lanes, or until no vehicle could pass because the served lanes were blocked.
The next phase was the one whose queue had reached the upstream junction, if any, otherwise the next phase in the cycle that had waiting vehicles and had not yet been served in the current cycle.

\subsection{Experimental Scenarios and Performance Metrics}

Three sets of experiments addressed our research questions. Every network configuration was simulated over the full factorial sweep of demand levels $\lambda$ and surge amplitudes $a$, with 10 replications for each combination.

We began with the five traffic management strategies applied to the whole network, so that all 16 intersections were either converted to roundabouts or signalized with fixed-time, actuated, Analytic+, or AcoLight methods.
Partial conversion came next.
Here, the 16 intersections of each network were ranked by importance, and the top 1, 4, 8, 12, and 16 were converted to roundabouts while the rest continued with actuated control, which made a network with no conversions identical to the all-actuated baseline.

Two rankings were compared.
Unweighted betweenness centrality counts how often an intersection lies on the shortest path between other pairs of intersections, and is therefore a purely structural measure.
Intersection demand, its weighted-degree counterpart for traffic networks \citep{Barrat2004Weighted}, instead scores each intersection by the total traffic density on its incident edges and thereby reflects where traffic concentrates under the simulated demand pattern.

For each vehicle $v$, SUMO records its travel time from insertion to arrival, its route length, and its time loss
\begin{equation}
    \ell_v=\int \left(1-\frac{u_v(t)}{\hat{u}_v(t)}\right)\mathrm{d}t,
    \label{eq:timeloss}
\end{equation}
where $u_v$ is its speed and $\hat{u}_v$ is the speed limit.
$\ell_v$ is the extra travel time compared with free-flow travel along the same route, so a vehicle driving at the limit has $\ell_v=0$.
From these records we obtain the mean trip time, the travel time plus the time $\delta_v$ that vehicle $v$ waited after its scheduled departure before it could be inserted into the network, and the mean speed, the mean route length divided by the mean travel time.

Trip times and speeds depend on route length and on the speed limit as well as on the traffic management strategy.
Time loss is an absolute measure of the additional travel time incurred over free-flow travel time.
We also use delays as a relative (in)efficiency metric to compare traffic management strategies fairly across routes of different length and networks with different limits.

We combine time loss with the time spent outside the saturated network to get the mean vehicle delay
\begin{equation}
    D=\frac{1}{N}\sum_{v=1}^{N}\left(\ell_v+\delta_v\right).
    \label{eq:vehdelay}
\end{equation}
This vehicle-based (in)efficiency measure is defined only for the $N$ vehicles that complete their trip.
We use this to compare the whole-network strategies (Sec.~\ref{sec:full}) and for the speed-limit comparison (Sec.~\ref{sec:speed}), together with the mean trip time and the mean speed.

The network delay is an edge-based measure of the time loss. For edge $e$ and measurement interval $k$ of length $\Delta = 300$\,s, it is
\begin{equation}
    d_{e,k}=\Delta\,\frac{L_{e,k}}{S_{e,k}},
    \label{eq:netdelay}
\end{equation}
where $L_{e,k}$ is the total time loss accumulated on the edge during the interval and $S_{e,k}$ is the total time vehicles spent on it, so that $S_{e,k}/\Delta$ is the mean number of vehicles present; $d_{e,k}$ is thus the time lost per vehicle present, capped at $\Delta$.
The network delay is the unweighted mean of $d_{e,k}$ over all edges and intervals with vehicles present, excluding the ring edges of converted junctions.
Unlike the mean vehicle delay, it omits the time vehicles spend queued outside the network, but its edge-level record is needed for the time-resolved response to surges (Sec.~\ref{sec:full}) and for partial conversion (Sec.~\ref{sec:partial}).

\section{Results}
\label{sec:results}

\subsection{Full Conversion against Signal Control}
\label{sec:full}

\begin{figure}[tb]
    \centering
    \includegraphics{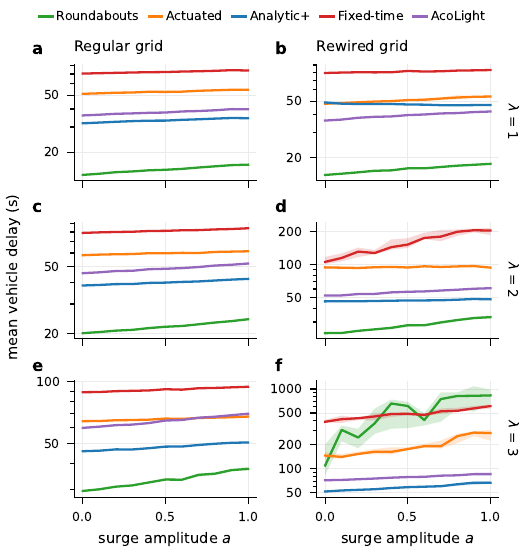}
    \caption{Median of the mean vehicle delay against surge amplitude for the five whole-network strategies on the regular grid (a, c, e) and the rewired grid (b, d, f) for different  $\lambda$.
    Bands show the interquartile range of 10 seeds.}
    \label{fig:degradation}
\end{figure}

Demand on both networks follows a square wave. For all strategies, delay increases during each surge half-period and decreases during recovery (see Supplementary Fig.~\SuppRef{fig:oscillation}). 
We compare strategies using mean vehicle delay per run, taking medians across 10 replications. 
On the regular grid, the fully converted network yields the lowest delay across all demand levels and surge amplitudes tested (Fig.~\ref{fig:degradation}a,c,e). 
Among the management strategies, self-organizing signals perform best, outperforming actuated control up to 2$\times$ base demand and matching it at 3$\times$. 
Fixed-time is worst, while Analytic+ is the best signal control strategy, especially at base demand.

In the irregular topology, the self-organizing signals increasingly outperform actuated control as the load grows.
At the base demand ($\lambda=1)$, actuated control and Analytic+ are almost indistinguishable, while AcoLight is the best of the management strategies (Fig.~\ref{fig:degradation}b). 
At higher demands, Analytic+ outperforms AcoLight (Fig.~\ref{fig:degradation}d,f).

Surges affect the self-organizing signals least: across the amplitude sweep at that demand level, delay rises only to 67 and 86\,s. 
Fixed-time control, by contrast, reaches 611\,s, with only 74\% of vehicles completing trips. 
Converting all signalized intersections to roundabouts performs best of the five strategies, up to twice base demand at every surge amplitude.

On the rewired network, however, the advantage of roundabouts ends abruptly for high demands.
At $\lambda = 3$, the mean vehicle delay jumps from 110\,s under steady demand to several hundred seconds when surges are present, and only 29--65\% of loaded vehicles complete their trips. 
The network delay grows over the run without stabilization (Fig.~\ref{fig:degradation}f; Supplementary Fig.~\SuppRef{fig:oscillation}), while the performance of signalized networks degrades only slowly. 
In the regular grid instead, a full conversion of intersections into roundabouts delivers a superior performance even at $\lambda = 3$ and in presence of surges ($a = 1$), with less network delays than produced by any management strategy studied.

\begin{figure}[tb]
    \centering
    \includegraphics{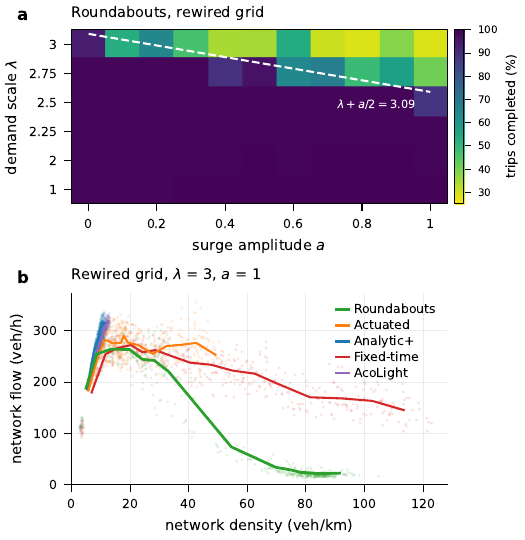}
    \caption{Breakdown of the fully converted rewired network.
    (a) Median share of loaded vehicles completing their trips, 10 seeds per cell; the dashed line, $\lambda + a/2 = 3.09$, follows constant time-averaged demand, whereas a boundary set by peak demand $\lambda + a$ would fall twice as steeply.
    (b) Macroscopic fundamental diagram at $\lambda = 3$, $a = 1$: network-average flow against density, one point per 5-minute interval and seed, with binned means.}
    \label{fig:breakdown}
\end{figure}

The breakdown of fully converted networks is determined by time-averaged rather than peak demand (Fig.~\ref{fig:breakdown}a).
Taking its onset at each surge amplitude as the first demand level at which fewer than 95\% of loaded vehicles complete their trips, the onset moves from $\lambda = 3$ at small amplitudes to $\lambda = 2.5$ at $a = 1$, but the time-averaged demand $\lambda + a/2$ there stays at $3.09 \pm 0.09$.
Past that load, congestion sets in and flows drop, as also reflected by the macroscopic fundamental diagram~\cite{Geroliminis2008MFD, Daganzo2008MFD} (Fig.~\ref{fig:breakdown}b).
Although actuated and fixed-time control also cause congestion, the flows in the fully converted network drop much more drastically.
In contrast, the self-organizing signals and every strategy on the regular grid stay on the free-flow branch.

Below this boundary, the fully converted network is also resilient to volatility.
A system is fragile if a fluctuating load costs more than a constant load carrying the same number of vehicles \citep{Taleb2012Antifragile}, which we test by subtracting from each oscillating run's mean vehicle delay that of a steady run at the same mean demand $\lambda + a/2$ (Supplementary Fig.~\SuppRef{fig:convexity}).
Under full conversion, this gap stays within 1\,s of zero at $\lambda = 1$ and 2 on both networks, as it does for every signal on the regular grid and for the self-organizing signals on the fully rewired one.
At $\lambda = 3$, on the rewired network it reaches 525\,s under full conversion against 7\,s for the self-organizing signals, while on the regular grid it stays below 10\,s.
These values are upper bounds, since the steady reference cannot be extended beyond $\lambda = 3$.

\subsection{Partial Conversion and Node Selection}
\label{sec:partial}

\begin{figure}[tb]
    \centering
    \includegraphics{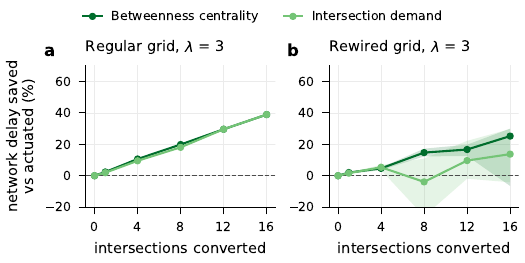}
    \caption{Network-delay reduction (higher is better) relative to actuated control against the number of converted intersections, for the two selection rules under steady demand at $\lambda = 3$ on (a) the regular and (b) the rewired grid.
    Zero conversions is the actuated baseline (dashed); betweenness centrality selects intersections better than intersection demand.}
    \label{fig:rules}
\end{figure}

On the regular grid, each conversion reduces delays at all demand levels (Fig.~\ref{fig:rules}a; Supplementary Fig.~\SuppRef{fig:selectionfull}). 
Relative to actuated control on the same network, demand level, surge amplitude, and seed, the median steady-demand delay reduction increases from 3\% with one converted intersection to 62\% with sixteen, with surges having little effect. 
Converting only the most central intersection cuts delays by 2--3\% across all tested demand levels and surge amplitudes. 
At the highest demand, full conversion is still 15--23\% better than the best signalized alternative (Analytic+).

For the irregular topology under steady demand, centrality-ranked conversions yield increasing benefits across all demand levels (Fig.~\ref{fig:rules}b). 
At triple demand ($\lambda=3$), delays drop by 2\% with one conversion and by 15\% with eight. 
Under the heaviest demand with surges, this trend fails.
Here, twelve conversions cut delays by 17\% in steady conditions but become 11–49\% worse than actuated control with any surge, mirroring the fully converted network. 
Networks with one or four conversions avoid breakdown and stay within 5\% of actuated control; under the most stressful conditions, eight conversions perform best, though gains diminish as surges grow. 
At $\lambda=3$ and $a=1$, the centrality-ranked 8-node network has a median delay of 114\,s, which is better than full conversion (164\,s) and actuated control, but worse than Analytic+ (97\,s).

We find that node-selection strategy has little effect across demand levels (Fig.~\ref{fig:rules}a), except on the rewired grid at high demand (Fig.~\ref{fig:rules}b). 
With eight converted intersections under steady demand, centrality-based selection reduces delays by 15\% versus actuated control, while demand-based selection increases delays by 4\% on average (with mixed results across the 10 seeds).
The two selections share six junctions and both yield connected subnetworks, so the difference hinges on which two remaining junctions are converted. 
Centrality-based selection is consistent across surge amplitudes, whereas the demand-based network performs 25--50\% worse once any surge occurs (Supplementary Fig.~\SuppRef{fig:robustness}).

Thus, the answer to how intersections should be converted is conditional.
With steady demand, hybrid networks (signalized intersections plus roundabouts) improve as more intersections are converted in both topologies. 
In the irregular network under heavy surges, however, many conversions cause breakdown, so the optimal number and choice of intersections depends on load. 
The two selection rules are equivalent except at eight conversions in the irregular network under the heaviest demand.
There, centrality-based ranking matches or outperforms actuated control across the surge range, while demand-based ranking performs much worse.

\subsection{Is Slower Faster? Speed-Limited Roundabouts}
\label{sec:speed}

\begin{figure}[tb]
    \centering
    \includegraphics{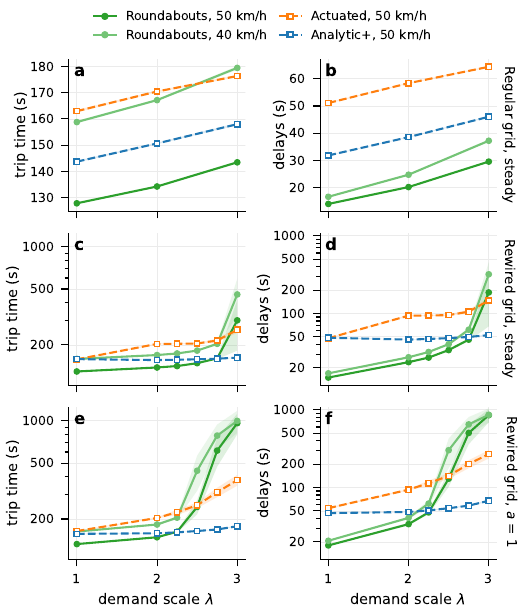}
    \caption{Fully converted roundabout networks at 50 and 40\,km/h vs. signalized intersection network at 50\,km/h: mean trip time (a, c, e) and delays $D$ (b, d, f; Eq.~\eqref{eq:vehdelay}) on the regular grid under steady demand (a, b), where surges leave the ordering unchanged, and on the rewired grid under steady demand (c, d) and at $a = 1$ (e, f).
    Means and 95\% confidence intervals over 10 seeds.
    For the other traffic management strategies, see Supplementary Fig.~\SuppRef{fig:speedfull}.}
    \label{fig:speedcomparison}
\end{figure}

In the following, we will show that a network-wide speed reduction, adopted for safety, noise or environmental reasons, does not necessarily raise travel times.
We demonstrate this slower-is-faster effect~\cite{Gershenson2015SlowerIsFaster} for mean trip times and for delays $D$ (Fig.~\ref{fig:speedcomparison}).
Delays exclude the free-flow travel time, which is fixed at a certain value, so it is the part of travel time that varies with traffic.

For a speed limit of 40\,km/h, the converted regular grid has shorter trip times than actuated control for a speed limit of 50\,km/h, unless demand is very high, while Analytic+ at 50\,km/h has shorter trip times throughout (Fig.~\ref{fig:speedcomparison}a).
However, the delays for roundabouts at 40\,km/h are still lower than under \textit{any} management strategy at 50\,km/h, in every condition tested (Fig.~\ref{fig:speedcomparison}b).
The two rankings differ, because a lower speed limit lengthens every trip time from the outset.
Trips through the 40\,km/h roundabouts therefore take longer than under Analytic+ at 50\,km/h. Yet, they deviate less from their ideal, undisturbed duration.
The rewired network confirms this trend in the breakdown shown below (Fig.~\ref{fig:speedcomparison}c--f).
From twice the base demand onward, it pulls further ahead of actuated control, which shows higher delays than on the grid.

\begin{figure}[tb]
    \centering
    \includegraphics{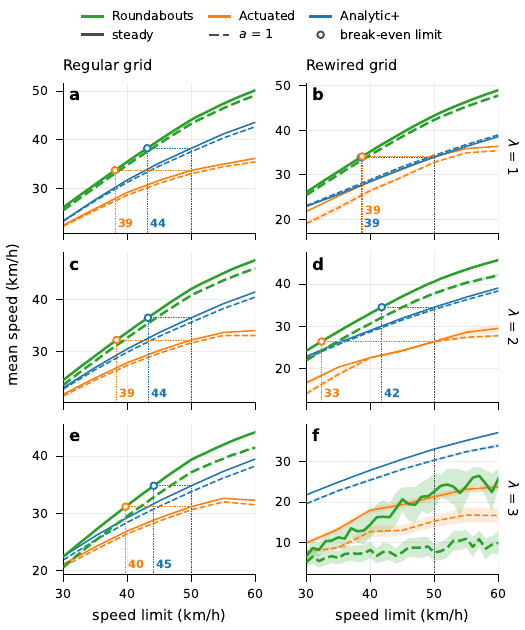}
    \caption{Mean speed vs. speed limit on the regular grid (a, c, e) and the rewired grid (b, d, f) under steady demand (solid) and with surges at $a = 1$ (dashed).
    The break-even limit denotes where the roundabout network matches the signals at 50\,km/h under steady demand.
    (f) At high demand, the converted network breaks down at every speed limit.
    Means and 95\% confidence intervals over 10 seeds.}
    \label{fig:speedlimit}
\end{figure}

Across the full range of speed limits (Fig.~\ref{fig:speedlimit}), down to a limit of about 40\,km/h, trips through the converted regular grid are at least as fast as under actuated control at 50\,km/h. 
They are as fast as under Analytic+ down to about 45\,km/h, nearly independently of demand level and surge amplitude (Fig.~\ref{fig:speedlimit}a,c,e). 
In comparison with fixed-time control, the speed limit in the converted network can be lowered even further.
The rewired network leaves at least as much room before breakdown (Fig.~\ref{fig:speedlimit}b,d):
At equal speed limits, the converted network is faster than both traffic signal strategies, and its lead grows with the speed limit. 
This is because waiting at red does not shorten as the speed limit rises and actuated control gains almost nothing above 50\,km/h; only under the heaviest loads Analytic+ match or outperform the roundabouts.

At three times the base demand the converted rewired network breaks down for every speed limit and the order reverses: Analytic+ is fastest for every limit speed, and actuated control is ahead of roundabouts at low speed limits, but under surges  at every speed limit  (Fig.~\ref{fig:speedlimit}f).
The speed limit shifts this boundary more readily downwards than upwards, since the lowest speed limits cause break downs at visibly lower demands, even more so under surges. Speed limits above 50\,km/h increase it only under steady demand; close to the boundary the outcome varies strongly between seeds and neighboring limits.

Within the converted network, every reduction of the speed limit lowers the mean speed at every demand level and amplitude, although not uniformly: below 50\,km/h, most of each additional km/h carries through to the mean speed, while above it vehicles benefit less and less from the higher speed limit on the short links between roundabouts.

Summarizing the above, we find a slower-is-faster effect in the sense that a road networked with (actuated) signalized intersections and a speed limit of 50\,km/h is typically outperformed by the same network at $\sim$40\,km/h, where intersections were converted into roundabouts.
Against Analytic+, the gain is a bit lower, where the break even point lies $\sim$45\,km/h.
The exception are rewired grids at very high demand levels, where the traffic flow in converted networks breaks down; to avoid this breakdown, traffic signal control is needed.

\section{Discussion}
\label{sec:discussion}
Intersection management is usually judged by its efficiency, through the delays it adds to trips or the travel times that results. Yet, efficiency is only one of the qualities a transport system should offer.
It should also perform reliably when demand fluctuates, generate predictable trip times, and avoid conflicts among vehicles.
On a regular grid and on the irregular one up to twice the base demand, the roundabout network combines these qualities with the lowest delays, and intersections managed with self-organizing signals come closest to it.

Unless the demand level (in rewired networks) is too large, the efficiency gain is considerable.
On the regular grid full conversion cuts the mean vehicle delay by a factor of 2.5--5 relative to fixed-time control and by 1.3--2.3 relative to the best self-organizing signals. 
This applies at all demand levels and surge amplitudes studied, because continuous, gap-based service avoids idle green times and inefficient stop-and-go cycles that phased signals impose.
This gain does not require converting every intersection at once: under steady demand, every additional centrality-ranked roundabout lowers delays
(even for a single roundabout among actuated signals). Thus, a mixed network implies no penalties.
The improved performance can also be used to lower the speed limit: a roundabout network with a speed limit 5 to 10\,km/h below that in a signalized one achieves trip completion with about the same travel times.
Across traffic management regimes, this implies a slower-is-faster effect that can be used to improve safety, noise levels, and environmental impact. 

The roundabout network is reliable and predictable.
Up to twice the base demand, delays due to an oscillating demand profile are worse than a steady load with the same mean for the management strategies in the rewired network.
Roundabouts also minimize trip-time variability \cite{Huberman1999Unstable}: on the regular grid, the standard deviation is about one-third of the mean for all strategies, so the fastest trips are also the most predictable; on the rewired network, roundabouts have the smallest spread up to twice the base demand (see Supplementary Fig.~\SuppRef{fig:reliability}).
A lower speed limit does not change this, since it adds a fixed free-flow time, while signal-based delays depends the arrival phase and on demand. 
The simulations of our converted regular grids show no collisions; on the rewired network, collisions drop sharply with more conversions and remain low even when large conversions fail due to traffic breakdowns at large demand levels (see Supplementary Fig.~\SuppRef{fig:collisions}). This is consistent with fewer conflict points and lower approach speeds \cite{Rodegerdts2010NCHRP672, Persaud2001Safety}. 
AcoLight also records almost no simulated collisions, while most other traffic signal control strategies see collisions rise with demand and surges, so while travel times deteriorate with demand, safety is reduced as well. 

In our simulations, the advantages of roundabout-based rewired networks disappear once time-averaged demand reaches ~3.1$\times$ the base load. Then, traffic
breakdown (``gridlock'') occurs as circulating flows saturate, yield-controlled entries cannot find gaps, and irregular topology (with heterogeneous degrees and link lengths) creates unbalanced entry demands that collapse entry capacity even at a single roundabout~\cite{Rodegerdts2010NCHRP672, Vlahos2008Capacity}. 
In contrast, signalized intersections can still handle traffic flows at such high demands.

When the network gets overloaded, predictability fails first.
Near the boundary to the gridlocked regime, the variance of roundabout trip times grows much faster than their mean.
Even where roundabouts and Analytic+ have equal mean trip times, roundabout trips are therefore already less predictable.
For self-organizing signals, in contrast, the standard deviation of trip times stays below half the mean at all loads.

In the high-demand regime, self-organizing signals are most robust; Analytic+ yields about half the delay of a fully converted network at steady demand of 3$\times$ the base load, and far less once surges cause a converted network to break down.
This aligns with Swiss authorities replacing peak-saturated roundabouts with traffic-responsive signals \cite{Aschwanden2026NZZ}; the results support doing so near the boundary, while below it roundabouts remain preferable.

Near that boundary, the performance of partially converted networks depends on how many and which intersections are converted.
On the irregular network with surges and three times the base demand, converting 12–16 intersections into roundabouts performed worse than actuated control, while converting eight high-betweenness intersections can match or beat it, likely because the remaining signals act like some perimeter control, thereby keeping the flow below the critical load. Node selection is definitely important: under surges, choosing intersections by betweenness centrality rather than demand could flip a loss into a gain, likely because centrality captures through-traffic paths and thus the core’s ability to absorb inflow, while demand reflects only current local concentrations. In highly surge- and demand-challenged networks, however, Analytic+ yields lower network delays. 

Having compared classical traffic signal control (based on fixed-time or actuated approaches) with two paradigms of self-organization (traffic flows in a roundabout-based network and in intersection-based networks with self-organizing signals), we can summarize our main findings as follows:
\begin{itemize}
    \item \textit{Self-organization vs. classical signal control:} Self-organized approaches to traffic flows in networks clearly beat classical signal control approaches. 
    \item \textit{Traffic organizing itself:} Without signals, roundabouts produce the lowest delays, insensitivity to surges, most predictable trip times, and fewest collisions, up to a critical load. Even at a lower speed limit, it may outperform signalized networks.
    \item \textit{Load as the deciding factor:} Irregular topology and surges bring a network to the critical load sooner. To avoid gridlock, traffic lights may be used. 
    \item \textit{Self-organizing signals helping traffic organize:} Self-organizing signals outperform classical traffic signal control (see also \cite{Bompard2026RuleBased}). For example, by responding to the actual local traffic flows, Analytic+ and AcoLight keep delays bounded and trip times predictable.
\end{itemize}
Obviously, by partial conversion, both paradigms may be combined. For example, perimeter control may be used to avoid an overload of converted parts of the traffic network~\cite{KeyvanEkbatani2012Gating, Geroliminis2013Perimeter}.
For instance, the City of Zurich, Switzerland, already does this at the edge of its inner city \cite{Ortigosa2014Zurich}.
A center protected in this way can keep its roundabouts below the critical load.

Despite its merits, our study also has some limitations. The networks are synthetic grids with a scaled demand pattern and fixed routes. Only cars are simulated, but no pedestrians, cyclists, or heavy vehicles. 
Lane use within roundabouts is simplified: although real drivers tend to use the inner lanes and merge out before exiting, SUMO applies right-of-way only within each junction, not between adjacent junctions on the ring, which created spurious collisions (see Supplementary Sec.~\SuppRef{app:safety}). 
Drivers therefore stay in the outer lane, likely underestimating multi-lane roundabout capacity and omitting lane-change conflicts that real designs reduce via lane separation (e.g., turbo roundabouts \cite{Fortuijn2009Turbo}).


\section{Conclusion}
\label{sec:conclusion}

This study studied spatially decentralized approaches to self-organized traffic flows in comparison with classical traffic signal control, thereby implicitly comparing bottom-up coordination with top-down control. 
Specifically, we simulated road networks based on roundabouts and signalized intersections. 
Conventional signal control (based on fixed-time or actuated approaches) was compared with self-organizing signals (Analytic+ and AcoLight). 
This was done on a 4$\times$4 grid and a rewired irregular network for demands up to 3$\times$ a base load and with periodic surges amounting up to 100\% of it.
Even at lower speed limits, roundabout networks could reach trip times matching signalized intersections at 50\,km/h and yielding lower delays (thereby implying a slower-is-faster effect). 

Hence, roundabout networks outperformed all signalized alternatives. 
They also absorbed large surges with no measurable penalty versus an equally loaded steady demand. 
However, this applied only up to a critical load. 
While in the regular grid this critical load was never reached in our simulations, in the irregular network, full conversion failed at about 3.1$\times$ the base demand. 
Then, self-organizing signals based on Analytic+ or AcoLight performed best. 
Partial conversion reduced delay under steady demand for any number of converted intersections; near capacity on the irregular network, it mattered both how many and which intersections were converted.
Node selection based on betweenness-centrality was more robust than demand-based selection. 

Overall, spatial and temporal decentralization were complementary: spatial decentralization (conversion of signalized intersections into roundabouts) works best for regular topologies below breakdown, while temporal conversion (replacing classical signal control by self-organizing signals) performed best for irregular networks under extreme load. 
We propose that future work should next compare our computer simulation results with real-world traffic flows.

\begin{small}
\bibliographystyle{IEEEtranN}
\bibliography{refs}
\end{small}




\ifarxiv
\clearpage
\onecolumn
\section*{Supplementary Material}

\renewcommand{\thesection}{S\arabic{section}}
\renewcommand{\thesubsection}{\thesection.\arabic{subsection}}
\renewcommand{\thefigure}{S\arabic{figure}}
\renewcommand{\thetable}{S\arabic{table}}
\renewcommand{\theequation}{S\arabic{equation}}
\setcounter{section}{0}
\setcounter{subsection}{0}
\setcounter{figure}{0}
\setcounter{table}{0}
\setcounter{equation}{0}

\section{Response to Oscillating Demand}
\label{app:supp}
The two figures below support Sec.~\MainRef{sec:full} of the main paper.
They show the time-resolved response to the oscillating demand and the cost of volatility at matched mean load.

\begin{figure*}[h!tb]
    \centering
    \includegraphics[width=0.8\textwidth]{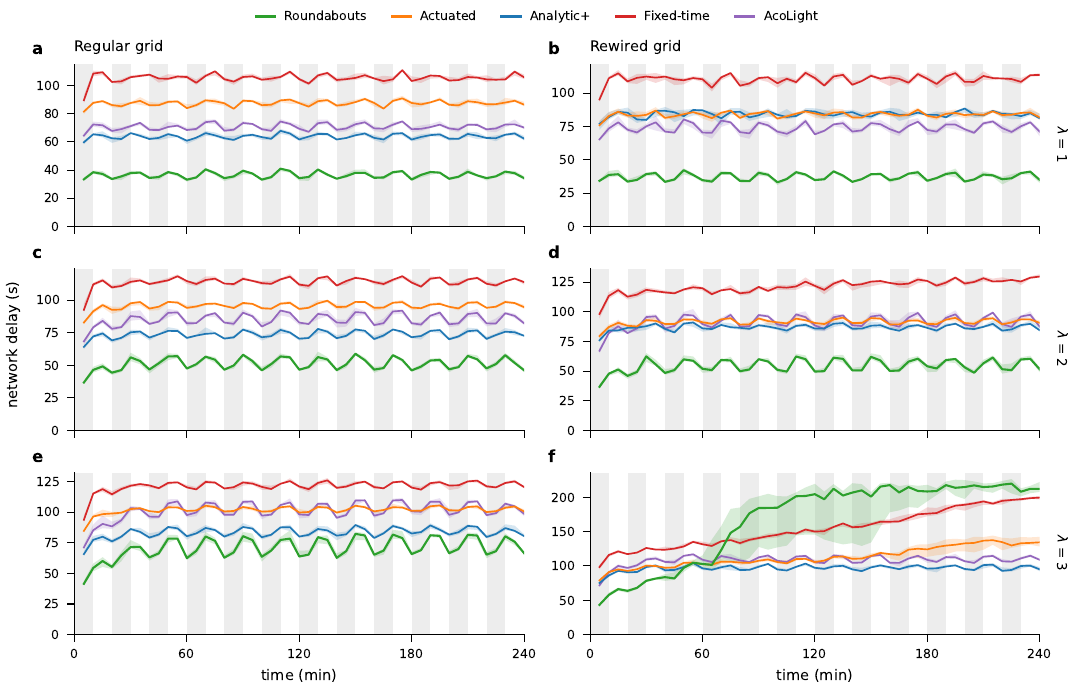}
    \caption{Mean network delay over the 4-hour horizon at $a = 1$, by network (columns) and demand level (rows).
    Gray bands mark the surged half-periods. Lines are medians, and colored bands the central 50\% of 10 seeds.}
    \label{fig:oscillation}
\end{figure*}

\begin{figure}[h!tb]
    \centering
    \includegraphics{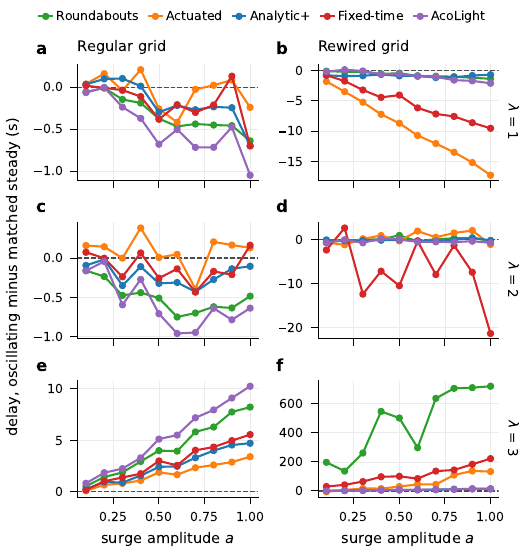}
    \caption{Cost of volatility: median of the mean vehicle delay minus that of a steady run at the same time-averaged demand $\lambda + a/2$, interpolated from the $a = 0$ runs, on the regular grid (a, c, e) and the rewired grid (b, d, f) at $\lambda = 1$, 2, and 3.
    Values at or below zero mean that oscillating demand costs no more than a constant load.
    At $\lambda = 3$ the steady reference is held at its $\lambda = 3$ value, so those gaps are upper bounds.
    On the rewired network, the negative gaps of actuated and fixed-time control reflect an overstated interpolated reference rather than a benefit.}
    \label{fig:convexity}
\end{figure}

\clearpage
\section{Macroscopic Network Dynamics}
\label{app:mfd}

\begin{figure*}[h!tb]
    \centering
    \includegraphics[width=\textwidth]{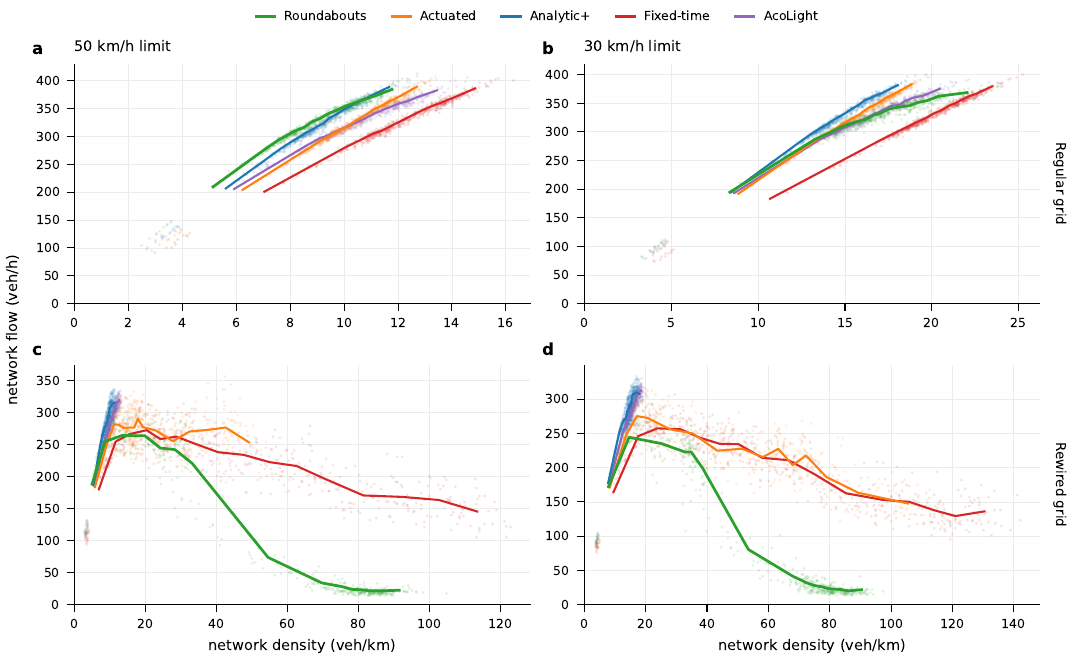}
    \caption{Macroscopic fundamental diagrams~\cite{Daganzo2008MFD} at $\lambda = 3$, $a = 1$: network-average flow against density, by network (rows) and speed limit (columns).
    Points are single runs in 5-minute intervals and lines binned means.}
    \label{fig:mfd}
\end{figure*}

The macroscopic fundamental diagram, network-average flow against network-average density (Fig.~\ref{fig:mfd}), characterizes the congested state behind the breakdown identified in Sec.~\MainRef{sec:full} of the main paper.
The condition shown is the most stressed one simulated, three times the base demand with the maximal surge amplitude, for the four signalized strategies and for the fully converted roundabout network at both speed limits.
On the regular grid every configuration stays on the free-flow branch, with flow rising as density rises and no capacity drop anywhere in the observed range.
The converted network operates on the steepest branch, with the highest network speed for a given accumulation, which is the macroscopic signature of its lower delays.
At accumulations of 10--15\,veh/km it sustains about 367\,veh/h, against 352\,veh/h for actuated and 333\,veh/h for fixed-time control.
Its binned mean flow peaks at about 384\,veh/h, level with the signals, at an accumulation of 12\,veh/km, against 13 to 15\,veh/km for all but Analytic+.
Its 30\,km/h counterpart peaks at a comparable 369\,veh/h, but at 22\,veh/km rather than 12\,veh/km, consistent with its lower free-flow speed.
Even under the heaviest load simulated, the regular grid never reaches the crossover demand anticipated in Sec.~\MainRef{sec:related} of the main paper, because the gap-acceptance capacity of the roundabouts is never exhausted.

On the rewired network, the converted network instead traces a sharply peaked diagram at both speed limits.
Its binned mean flow rises to about 260\,veh/h at accumulations near 13\,veh/km, similar to fixed-time control, and then collapses to about 20\,veh/h near 80\,veh/km.
That shape is the macroscopic signature of network gridlock, in which circulating rings block one another and entering vehicles can no longer find acceptable gaps.
Under these conditions the simulator teleports about 1{,}350 jammed vehicles per run in the fully converted network, against 263 under actuated control, a further indication that the breakdown is gridlock rather than a gradual loss of throughput.
The actuated and fixed-time signals degrade far more gracefully, holding binned mean flows between 129 and 276\,veh/h beyond 40\,veh/km and between 129 and 163\,veh/h at accumulations beyond 100\,veh/km.
Neither of the self-organizing signals enters that regime.
Under Analytic+, the accumulation stays below 13\,veh/km at the default limit and 21\,veh/km at the reduced one, and under AcoLight below about 14 and 20\,veh/km.
Both remain on the rising branch of the diagram, with flows of up to about 335\,veh/h.
The capacity boundary of spatially decentralized traffic management anticipated by the intersection-level literature~\cite{Vlahos2008Capacity, Rodegerdts2010NCHRP672} thus appears here at the network level.
Once the circulating streams saturate, the self-organized priority rule has no mechanism to break the deadlock, whereas signals can interrupt the flow, even though they are less efficient at low loads.

The actuated curve at the default limit spans a visibly narrower density range than that of fixed-time control, never exceeding about 61\,veh/km where fixed-time control reaches 122\,veh/km.
That narrow range reflects higher throughput rather than a truncated observation window.
Actuated control completes 17{,}307 trips per run on average, against 14{,}016 for fixed-time control and 7{,}444 for full conversion, and Analytic+ and AcoLight, whose ranges are narrower still, complete 18{,}530 and 18{,}508.
Low accumulation combined with high completion is thus the signature of a network that is clearing its load.

\section{Node-Selection Rules at All Demand Levels}
\label{app:selectionfull}

\begin{figure}[h!tb]
    \centering
    \includegraphics{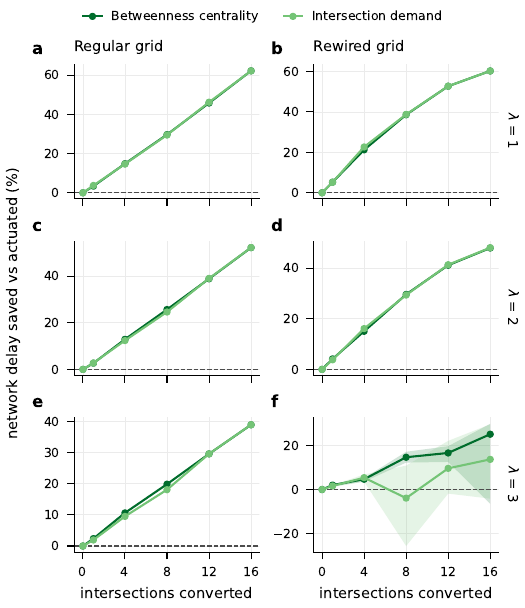}
    \caption{As Fig.~\MainRef{fig:rules} of the main paper at all three demand levels: network-delay reduction relative to actuated control against the number of converted intersections, for the two selection rules under steady demand, on the regular grid (a, c, e) and the rewired grid (b, d, f) at $\lambda = 1$, 2, and 3.
    Zero conversions is the actuated baseline (dashed); medians and central 50\% of 10 seeds.}
    \label{fig:selectionfull}
\end{figure}

Under steady demand, the reduction grows with every additional conversion on both networks and at every demand level.
The two selection rules give coinciding curves everywhere except on the rewired grid at $\lambda = 3$ (Fig.~\ref{fig:selectionfull}f), the case shown in the main text.

\clearpage

\section{Robustness of the Node-Selection Rules to Surging}
\label{app:robustness}

\begin{figure}[h!tb]
    \centering
    \includegraphics{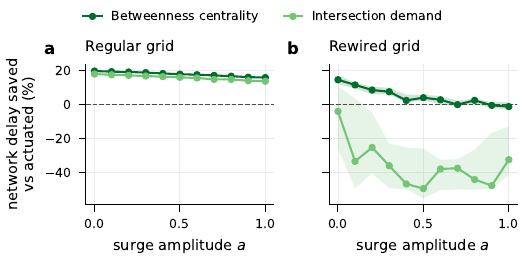}
    \caption{Network-delay reduction relative to actuated control against surge amplitude for the two selection rules, at eight conversions and $\lambda = 3$, on the regular grid (a) and the rewired grid (b).
    Medians and central 50\% of 10 seeds.}
    \label{fig:robustness}
\end{figure}

Holding the number of converted intersections at eight and the demand level at $\lambda = 3$, the centrality-ranked gain decays from 14.7\% under steady demand to $-1.1$\% at the maximal amplitude.
It is positive at nine of the eleven amplitudes and within about 1\% of actuated control at the other two.
The demand-ranked network, in contrast, goes from 3.9\% worse than actuated control under steady demand to between 25 and 50\% worse at every nonzero amplitude (Fig.~\ref{fig:robustness}).
On the regular grid both rules give a gain of 14--20\% at every amplitude, and the gap between them never exceeds 2.3 percentage points.

\section{Speed-Limited Roundabouts against All Management Strategies}
\label{app:speedfull}

\begin{figure}[h!tb]
    \centering
    \includegraphics{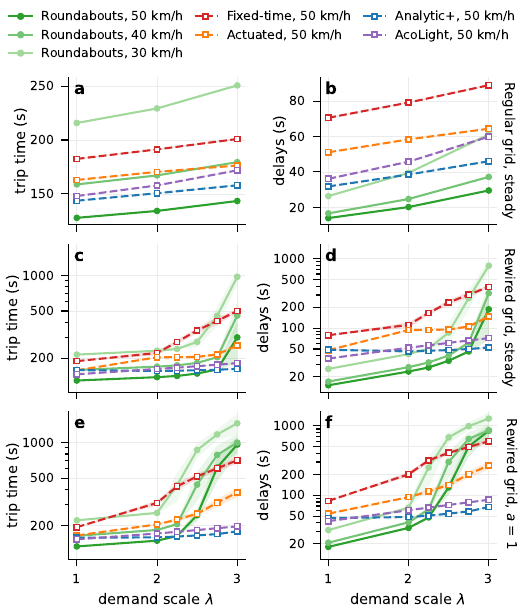}
    \caption{As the speed-limit comparison of Sec.~\MainRef{sec:speed} of the main paper, with the 30\,km/h roundabout network, fixed-time control, and AcoLight added: mean trip time (a, c, e) and mean vehicle delay $D$ (b, d, f; measured against each network's own limit) on the regular grid under steady demand (a, b) and on the rewired grid under steady demand (c, d) and at $a = 1$ (e, f).
    Means and 95\% confidence intervals over 10 seeds; panels c--f share logarithmic axes.}
    \label{fig:speedfull}
\end{figure}

Fig.~\ref{fig:speedfull} adds the configurations left out of the main-text comparison.
Fixed-time control is the slowest of the signals throughout, so the 40\,km/h roundabout network, which has shorter trips than actuated control up to twice the base demand, is ahead of it by a wider margin.
AcoLight lies between Analytic+ and actuated control on both measures, except at the base demand on the rewired network, where it is the fastest strategy.
At 30\,km/h the roundabout network still has lower delays than actuated and fixed-time control up to twice the base demand, but the free-flow cost of the limit exceeds that saving, and on the regular grid its trips are longer than under every management strategy.
On the rewired network it is also the first to break down, from $\lambda = 2.25$ under surges.

\section{Travel-Time Reliability}
\label{app:reliability}

\begin{figure}[h!tb]
    \centering
    \includegraphics{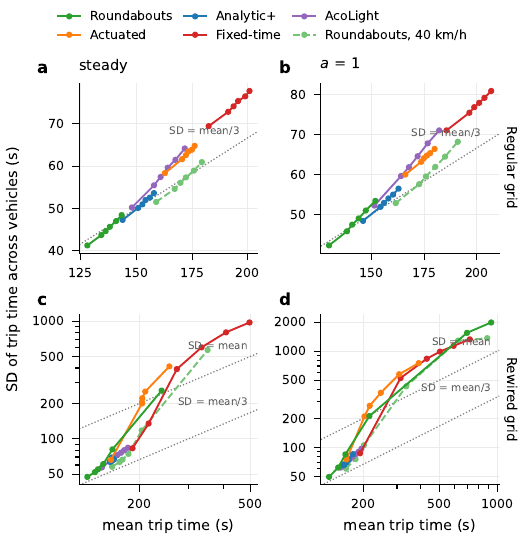}
    \caption{Spread of trip times between vehicles against their mean, after \citet{Huberman1999Unstable}: standard deviation against mean of the trip time, including the wait to enter the network, one point per demand level from $\lambda = 1$ to 3 (medians over 10 seeds), under steady demand (a, c) and at $a = 1$ (b, d), on the regular grid (a, b, linear axes) and the rewired grid (c, d, logarithmic axes).
    Dotted lines mark a standard deviation of a third of the mean and equal to the mean.
    Vehicles still in the network at the end are left out, so past breakdown the spread is a lower bound.}
    \label{fig:reliability}
\end{figure}

Fig.~\ref{fig:reliability} supports the discussion of predictability in Sec.~\MainRef{sec:discussion} of the main paper.
The trip times come from a repeat of the whole-network runs with per-vehicle trip records enabled, at $\lambda = 1$, 2, 2.25, 2.5, 2.75, and 3, under steady demand and at $a = 1$.
On the regular grid, every strategy lies close to the line on which the spread is a third of the mean, the roundabouts closest and the signals slightly above it.
A strategy with shorter trips thus also has more predictable ones (Fig.~\ref{fig:reliability}a,b).
On the rewired grid, the converted network, actuated control, and fixed-time control all leave that line as the load rises.
Once they have left it, they follow much the same curve, so how unpredictable trips become depends on how congested the network is rather than on the strategy that congests it (Fig.~\ref{fig:reliability}c,d).
Analytic+ and AcoLight stay close to the line at every load.

\clearpage

\section{Simulated Safety-Critical Events}
\label{app:safety}
This section complements the efficiency results of Secs.~\MainRef{sec:full} and~\MainRef{sec:partial} of the main paper with SUMO's safety proxies.
As crash data do not exist for a simulated network, we use proxy measures, which quantify conflict exposure rather than crash frequency~\cite{Gettman2003SSM}.
For safety, we count three events reported by SUMO: collisions, including conflicts inside junctions; emergency stops, in which a vehicle halts abruptly to resolve an imminent conflict; and emergency braking, in which a vehicle decelerates more strongly than its configured maximum deceleration.

\begin{figure}[h!tb]
    \centering
    \includegraphics{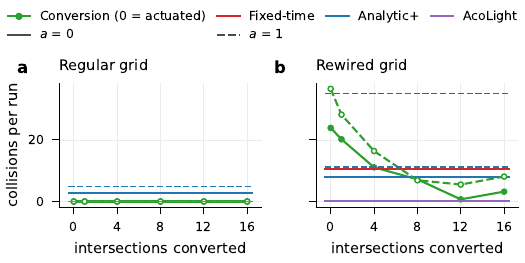}
    \caption{Mean collisions per run at $\lambda = 3$ against the number of centrality-ranked conversions (0 = all actuated), under steady demand (solid) and at $a = 1$ (dashed); horizontal lines mark fixed-time control, Analytic+, and AcoLight; (a) regular grid, (b) rewired grid.
    Emergency stops are zero and emergency braking rare in every configuration, so collisions dominate the safety-critical events.}
    \label{fig:collisions}
\end{figure}

Converting a substantial share of the network nearly eliminates simulated safety-critical events, in the direction the empirical before-and-after literature predicts (Fig.~\ref{fig:collisions}).
Of the three proxy events counted, emergency stops are zero in every configuration of both networks and emergency braking events are rare, so the figure plots collisions alone.
On the regular grid every centrality-ranked conversion, partial or full, records zero collisions at every condition tested (Fig.~\ref{fig:collisions}a).
On the rewired network at three times the base demand (Fig.~\ref{fig:collisions}b), any conversion of eight or more intersections records fewer collisions than fixed-time, actuated, and Analytic+ strategies at every surge amplitude, at most 8.5 per run against 11 to 37 for those signals at the maximal amplitude.
This is consistent with the reduction of conflict points and approach speeds documented in the empirical literature.
A four-way signalized intersection has 32 vehicle-to-vehicle conflict points, including crossing and left-turn conflicts, whereas a single-lane roundabout has 8 merging or diverging points~\cite{Rodegerdts2010NCHRP672, Persaud2001Safety}.
AcoLight is the exception among the signals, with at most 0.3 collisions per run in any condition, so signals can also match the safety record of conversion.
Collisions also increase with demand level and surge amplitude for most strategies that generate them in appreciable numbers, so safety and congestion stress are coupled.
The eight-node conversion, which stays near 7 per run at every amplitude, is the exception.

Apart from a single collision in 330 AcoLight runs, Analytic+ is the only configuration that records collisions on the regular grid, 2.7 per run at three times the base demand under steady demand and 4.9 at the maximal surge amplitude.
These residual collisions can be reduced further by lengthening the all-red interval, but only at the cost of network delay, since every additional second of all-red is time in which no movement is served.
The 2\,s interval used here is therefore a compromise between safety and delay rather than a safety optimum.
On the rewired network at that demand level it records fewer collisions than fixed-time and actuated control, 11.1 per run at the maximal amplitude against 36.6 for actuated control.

Sparse conversion on the rewired network is the least effective conversion on this measure, without being less safe than the signals it replaces.
At three times the base demand under steady demand the one-node conversion records 20.2 collisions per run, close to the all-actuated network from which it differs by a single junction (23.9).
Each further increment up to twelve lowers the count, to 0.7 at twelve converted intersections, and full conversion records a slightly higher 3.2.
The safety benefit therefore grows with the number of converted intersections up to twelve, and unlike the delay benefit it persists in the large conversions that break down under surges, where the fully converted network still records only 8.1 collisions per run.
These are proxy counts produced by deterministic, homogeneous simulated drivers, so they support comparisons between configurations rather than absolute statements.

The converted junctions keep the ring connections that SUMO generates, in which only the outer ring lane reaches the exits, and vehicles are set not to move to the inner ring lanes (\texttt{lcCooperativeRoundabout} $= 0$), so circulating traffic in effect uses a single lane.
Assigning lanes as in multi-lane roundabout practice, with the outer lane reserved for exiting traffic and through and left-turning vehicles on the inner lanes, would let the ring carry more traffic, but it also adds conflict points where vehicles change lanes to leave the ring or cross the path of entering traffic.
SUMO resolves conflicts only among the internal lanes of a single junction, while the ring edges between junctions are ordinary lanes on which no right of way is evaluated, so on closely spaced junctions a vehicle on a junction's internal lane can conflict with traffic at the neighboring junction without either yielding.
In a pilot with such lane assignments, these unresolved conflicts produced collisions at the junctions of the rewired grid where diagonal approaches meet the ring, although none on the regular grid.
Allowing vehicles onto the inner lanes of the generated rings raised collisions to hundreds per run on the rewired grid, and on the regular grid under surges, where it also caused gridlock.
We therefore retain the generated single-lane operation, which understates the capacity of the multi-lane rings and is in this respect conservative for the roundabout results.

\fi

\end{document}